\documentclass[conference]{IEEEtran}
\IEEEoverridecommandlockouts
\usepackage{caption,subcaption,cite}
\usepackage{amsmath,amssymb,amsfonts}
\usepackage{graphicx}
\usepackage{textcomp,makecell,multirow}
\usepackage[table]{xcolor}
\usepackage{booktabs,svg,pgfplots,tikz,tikzscale,adjustbox}
\usetikzlibrary{calc,positioning,patterns}
\usepgfplotslibrary{external} 
\pgfplotsset{compat=newest}
\newcounter{colend}
\newcounter{maxcol}
\newcounter{mincol}

\newcounter{facnt}
\newcounter{hacnt}
\newcounter{dotcnt}

\newcommand*\drawdotcol[3][0]{
  \setcounter{colend}{#3}
  \ifnum \value{colend} > 0 {   
    \addtocounter{colend}{-1}%
    \addtocounter{colend}{#1}%
    \foreach \t in {#1,...,\value{colend}}{
      \draw[fill=black] (\value{maxcol}-#2,\t) circle (0.2);
      \addtocounter{dotcnt}{1}
    }
  }
  \else {
    \draw[opacity=0,fill=black] (\value{maxcol}-#2,0) circle (0.2);
  }
  \fi
}

\newcommand*\drawtrunccol[3][0]{ 
  \setcounter{colend}{#3}
  \addtocounter{colend}{-1}%
  \addtocounter{colend}{#1}%

  \ifnum #3 > 0   
    \foreach \t in {#1,...,\value{colend}}{
      \draw[black!50, fill=white] (\value{maxcol}-#2,\t) circle (0.2);
  	  \addtocounter{dotcnt}{1}
    }
  \fi
}

\usepackage{rotating}
\usepackage[hidelinks]{hyperref}
\usepackage{hyperref}

\usepackage{balance}

\usepackage{multirow}
\usepackage{array}
\newcolumntype{C}[1]{>{\centering\arraybackslash}p{#1}} 
\newcolumntype{R}[1]{>{\raggedleft\arraybackslash}p{#1}} 
\newcolumntype{L}[1]{>{\raggedright\arraybackslash}p{#1}} 
\newcommand*\circled[1]{\tikz[baseline=(char.base)]{\node[shape=circle,draw,inner sep=0.5pt] (char) {#1};}}

\begin{document}
\title{Multiplier Design Addressing Area-Delay Trade-offs by using DSP and Logic resources on FPGAs}

\author{\IEEEauthorblockN{Andreas Böttcher, Martin Kumm}
\IEEEauthorblockA{\textit{Fulda, University of Applied Sciences, Faculty of Applied Computer Science}, Fulda, Germany \\
andreas.boettcher@cs.hs-fulda.de, martin.kumm@cs.hs-fulda.de}
}

\maketitle

\begin{abstract}
The major challenge when designing multipliers for FPGAs is to address several trade-offs:
On the one hand at the performance level and on the other hand at the resource level utilizing DSP blocks or look-up tables (LUTs).
With DSPs being a relatively limited resource, the problem of under- or over-utilization of DSPs has previously been addressed by the concept of multiplier tiling, by assembling multipliers from DSPs and small supplemental LUT multipliers. 
But there had always been an efficiency gap between 
tiling-based multipliers and radix-4 Booth-Arrays. 
While the monolithic Booth-Array was shown to be considerably more efficient in terms of LUT-resources on many modern FPGA-architectures, it typically possess a significantly higher critically path delay (or latency when pipelined) compared to multipliers designed by tiling. 
This work proposes and analyzes the use of smaller Booth-Arrays as sub-multipliers that are integrated into existing tiling-based methods, such that better trade-off points between area and delay can be reached while utilizing a user-specified number of DSP blocks.
It is shown by synthesis experiments, that the critical path delay compared to large Booth-Arrays can be reduced, while achieving significant reductions in LUT-resources compared to previous tiling.
\end{abstract}


\section{Introduction}
With multiplication being one of the prevalent operation in computer arithmetic its efficient and fast implementation is of ongoing interest. Contrary to application-specific integrated circuits (ASICs) the circuit design for FPGAs is stronger constrained by the predefined hardware architecture, with look-up-table (LUT) resources and embedded DSP blocks being the most relevant building blocks of arithmetic circuits. The embedded DSPs of modern FPGAs are the easiest applicable resource when implementing multiplication as among other operations they can directly compute a multiplication (25$\times$18 signed or 24$\times$17 unsigned on AMD Ultrascale(+) or up to a single signed or unsigned  27 $\times$27 or two 18$\times$18 on Intel Stratix 10). But as they are only available in relatively limited quantities, when solely relying on them for implementing multiplication, they can quickly become the limiting factor for implementing complex circuits \cite{Dinechin09,Banescu10,Walters14,Kumm15,Walters16,Ullah21,Kumm17}. Hence is of vital interest for multiplier design strategies to account for their efficient utilization and a matter of ongoing research\cite{Dinechin09,Banescu10,ParandehAfshar11,Brunie13,fwsakw17,Kumm17,Langhammer19DSP,Boettcher20}. With LUTs being the most abundant and versatile resource on FPGAs, various methods have been proposed to offset some or all of the complexity for multiplication by additional logic-based circuity to overall use less DSPs for an equivalent operation \cite{Dinechin09,Banescu10,ParandehAfshar11,Brunie13,Walters14,Kumm15,Walters16,fwsakw17,Kumm17,Langhammer19,Langhammer19DSP,Langhammer20DSP,Boettcher20,Ullah21}. 

The quest to utilize fewer DSPs within the design can be approached by different measures. One possibility is to apply mathematical substitutions like Karatsuba's algorithm to overall reduce the complexity of multiplication and thus use less DSPs, which had been demonstrated to also work on rectangular multipliers as on recent AMD FPGAs \cite{Dinechin09, Boettcher20}. Most approaches focus on achieving a higher DSP utilization and thus use less DSPs in total, as custom designs with application specific data width seldom utilize the maximal size of the embedded hardware multiplier. The extraction of multiple multiplications from a single DSP with auxiliary LUT-based circuity is proposed in \cite{fwsakw17,Langhammer19DSP,Langhammer20DSP,Xilinx2020}. For the same reason efficient LUT-based multipliers that utilize the Fast-Carry-Chain of modern FPGAs, like Bough-Wooley-Multipliers \cite{ParandehAfshar11}, Booth-Arrays \cite{Walters14, Kumm15, Walters16, Ullah21} or the Fractal-Synthesis LUT-mapping scheme \cite{Langhammer19DSP,Langhammer19}, are of great interest as an alternative to under-utilizing DSPs.

A systematic approach to assemble large multipliers from DSPs and additional LUT-based sub-multipliers was proposed with the multiplier tiling methodology \cite{Dinechin09}. It allows to weight between DSP- and logic-based resources by specifying limits for the maximal number and the minimal utilization of the DSPs used within a design and thus ensure an efficient use of DSP- and LUT-resources. The approach is agnostic to the type of sub-multiplier (called tiles), so each of the aforementioned multiplier design methods can themselves be used as a tile to generate larger multipliers. Several heuristic \cite{Banescu10,Brunie13,Boettcher20} and optimal \cite{Kumm17,Boettcher21} methods have been proposed to solve the tiling problem, such that for a given set of sub-multipliers and constraints, the solution requiring the least possible LUT-resources can be found.      
When assembling larger multipliers each sub-multiplier provides partial product bits which subsequently have to be added in a compressor tree to form the final product. As in contrast to ASICs a Dadda-Tree  \cite{Dadda65} with half- and full-adders does not map particularly efficiently on FPGAs \cite{bss93}, more complex structures that effectively use the 6-input LUTs and fast carry-chain present in most modern FPGA architectures like generalized parallel counters (GPCs)  \cite{Meo75,ParandehAfshar08,ParandehAfshar08b,ParandehAfshar09,ParandehAfshar11c,Kumm2014EfficientHS,Kumm14,Preusser17,Yuan19}, 4:2 row compressors \cite{Kumm2014EfficientHS,Hormigo09,Kamp09} and the ternary adder \cite{sp06,blsy09} were proposed. The multitude of additional compressors also requires compressor tree design methods which can accommodate for the placement of compressors with vastly different properties. Various FPGA-specific heuristic \cite{ParandehAfshar08,ParandehAfshar09,ParandehAfshar11c} and optimal \cite{ParandehAfshar08b,Matsunaga11,Matsunaga13,Kumm14,kk18} compressor tree design methods are available.

Additionally the aforementioned steps of partial-product-generator- and compressor-tree-design are not independent \cite{Boettcher23}, as a different selection of sub-multipliers can result in another combination of compressors being the combination with the least realization cost and vice-versa. So to represent the state-of-the-art and receive globally optimal results for a given set of sub-multiplier, compressors and final-adder-architectures a combined optimization as shown in \cite{Boettcher23} has to be applied.

In terms of LUT resources radix-4 Booth-Arrays \cite{Walters14,Kumm15,Walters16,Ullah21} are one of the most efficient concepts on FPGAs. They are a hierarchically structured array multiplier architecture, in which each level compresses the partial products from the previous to calculate the final result.
A recent proposal \cite{Nagar24} seeked to address the inherent limitation of Booth-Arrays that due to the stacking of logic levels, the critical path delay (or latency when pipelined) increases over-proportionally with size, compared to other multiplier design schemes. Thereby, a large multiplier was decomposed into several smaller Booth-Arrays, who's partial products were subsequently compressed with a Dadda tree. But Dadda trees do not represent the state-of-the-art on FPGAs, since compressor trees based on GPCs and row-adders typically utilize significantly less resources, have fewer stages and thus a lower critical path or latency. 
As previously mentioned, Booth-Arrays can be used as sub-multipliers with multiplier tiling. So when small Booth-Arrays are included in the available tile-set for the integer linear program (ILP)-model of \cite{Boettcher23}, equivalent sub-multiplier arrangements as in \cite{Nagar24} can be generated with tiling. This offers greater flexibility in term of the utilized sub-multipliers. Additionally DSPs can be included in the design and globally optimal solutions for tiling and the subsequent compression can be provided.

The present proposal %
\begin{itemize}
\item shows how Booth-Arrays can be used in conjunction with DSPs to form larger multipliers,
\item demonstrates the integration and use of Booth-Arrays with the multiplier tiling methodology to design medium and large multipliers,
\item investigates which level-depth of Booth-Arrays represents suitable compromise between resources and delay for tiling,
\item demonstrates that the proposed designs yield a reduction in utilized resources compared to other state-of-the-art logic-based multiplier design method while maintaining a similar critical path delay and latency.
\end{itemize}

\section{Multiplier Tiling}
\begin{figure}[t]
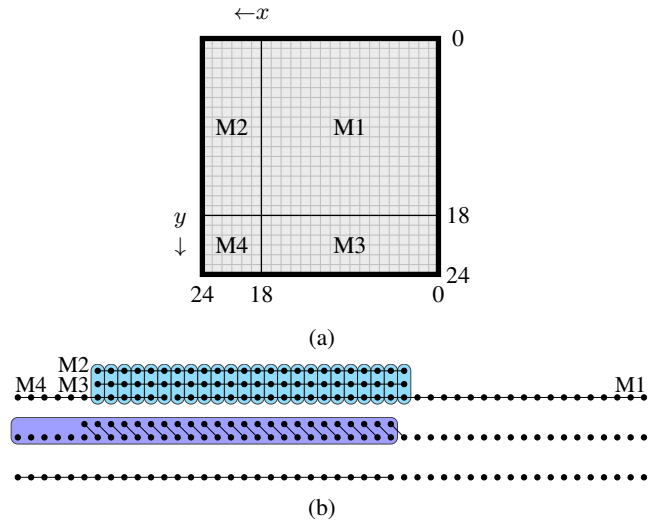

\centering
\begin{subfigure}[b]{0.23\textwidth}
    \centering
    \includegraphics[width=1\textwidth]{./tikz/6x4tiling_example.tikz}
    \caption{}
    \label{fig:6x6tiling}
\end{subfigure}

\begin{subfigure}[b]{0.48\textwidth}
    \centering
    \includegraphics[width=1\textwidth]{./tikz/6x4compression_example.tikz}
    \caption{}
    \label{fig:6x6compression}
\end{subfigure}
\caption{Tiling (a) and compressor tree (b) of a $24\times 24$-multiplier, composed of four sub-multiplier-tiles.}
\label{fig:multiplier_architecture}
\end{figure}

Imagine the design process for a 24$\times$24-multiplier as shown in \figurename~\ref{fig:multiplier_architecture} within the DSP-constrained context of a complex arithmetic circuit considering over- and under-utilization of DSPs to ensure their economic usage. Using a DSP block of modern Intel architectures with their full size of 27$\times$27 results in a certain under-utilization, which might motivate to use the DSP in its subdivided 18$\times$18 configuration (M1 in \figurename~\ref{fig:6x6tiling}) to retain the other halve for another multiplier and realize the rest of the 24$\times$24-multiplier with logic-based sub-multipliers ($M_{2,..,4}$). The large multiplier can be subdivided by decomposing the input vectors $X,Y$ into weighted MSB and LSB sub-words $x_h,x_l$ and $y_h,y_l$ as follows: 
\begin{align}
\label{eq:schoolbook_tiling}
X \times Y&=\left(x_h 2^{18}+x_l\right)\left(y_h 2^{18}+y_l\right) \notag \\
&=\underbrace{x_h y_h}_{M_4}2^{18+18}+\underbrace{x_h y_l}_{M_3}2^{18}+\underbrace{x_l y_h}_{M_2}2^{18}+\underbrace{x_l y_l}_{M_1} \
\end{align}
When expanding the brackets, the large multiplication is represented by four sub-multiplications $M_{1..4}$ weighted according to their Manhattan-distance relative to the origin in the tiling visualization of  \figurename~\ref{fig:6x6tiling}. The decomposition of large multipliers can be performed down to individual partial-product-bits as they would occur in the AND-array of a Bough-Wooley-Multiplier.
With that, multipliers with random shapes and sizes can be utilized in tiling, requiring only that a solution covering every position on the board of \figurename~\ref{fig:6x6tiling} can be found. 
While the tiling example in \figurename~\ref{fig:6x6tiling} is relatively trivial, when expanding the set of utilized tiles and requiring optimal solutions, multiplier tiling is known to be an NP-complete optimization problem \cite{Kumm17}. 
Each sub-multiplier will generate a partial product-vector, which subsequently has to be compressed on a compressor-tree as shown in \figurename~\ref{fig:6x6compression} to form a unified output vector. Practical designs commonly result in a higher number of bits to be compressed and accordingly more efficient compressors like GPCs are used.

\subsection{Utilized Multiplier Tile-Set}
Several characteristics of the sub-multipliers used in the tile-set that are relevant for multiplier tiling are listed in Table~\ref{table:tiles_properties}. First of all sub-multipliers feature a shape which defines the positions on the board in the graphical tiling representation which are covered be the particular tile. In the utilized tile-set all multipliers possess rectangular shapes, denoted by their $x\times y$-dimensions in the type column. Note that the small LUT-based tiles feature a fixed size, while the larger variants like the $2\times k$  \cite{ParandehAfshar11} can be generated with a variable size in one dimension dependent of a parameter $k$. Although non-rectangular multipliers like the Karatsuba structures with rectangular DSPs which feature gaps are possible \cite{Boettcher20}, they were not selected in this work for the tile-set as they are only beneficial for very large multipliers. Furthermore, derived from their shape, the tiles cover a certain area $A_t$, the number of positions (potential partial-products) on the tiling-board. Note that the LUT-based tiles (top of Table~\ref{table:tiles_properties}) mostly cover a considerably smaller area than the DSP block of AMD FPGAs used as a tile (bottom column). The sub-multiplier have realization costs in terms of utilized DSPs ($D^t$) and LUT $\text{cost}^\text{mult}_t$ resources. 
The total realization cost of a tile is defined as
\begin{equation}
    \text{cost}^\text{tile}_t=\text{cost}^\text{mult}_t+\text{cost}^\text{comp}_t .\ 
\end{equation}
It consists of the own realization cost of the tile $\text{cost}^\text{mult}_t$
as well as 
the average compression cost with the utilized set of compressors, which was determined to be about $0.65$ LUTs per bit \cite{Kumm17}, leading to
\begin{align}
  \text{cost}^\text{comp}_t \approx 0.65\cdot w_\text{out} \ .
\end{align}
Based on the quotient between the area covered by a particular tile $A_t$ and its total realization costs ($\text{cost}^\text{tile}_t$), a measure for the efficiency $E_t$ in terms of covered area per LUT can be calculated as follows:
\begin{equation}
    E_t=\frac{A_t}{\text{cost}^\text{tile}_t}
\end{equation}
Note that the total realization costs $\text{cost}^\text{tile}_t$ and efficiency $E_t$ are provided for reference and comparisons of compressors, while a combined optimization of the tiling and compressor tree \cite{Boettcher23} is used which minimizes the exact costs in terms of LUTs.

\begin{table}[t]
\caption{Properties of previous LUT- and DSP based sub-multipliers \cite{Boettcher20} 
targeting AMD FPGAs, where $\text{cost}^\text{mult}_e$ and $\text{cost}^\text{tile}_t$ are the number of LUT6 for the single multiplier and multiplier+compression, respectively
}
\centering
\begin{adjustbox}{max width=0.48\textwidth}
\setlength\tabcolsep{3pt}
\begin{tabular}{ l c c c c c l c}
\toprule
Type & $A_t$ & $\text{cost}^\text{mult}_t$ & $\text{cost}^\text{tile}_t$ & $w_\text{out}$ & $E_t$ & $D^t$ \\
\midrule
1$\times$1                   &   1   & 1     & 1.65  & 1             & 0.625 & 0\\
1$\times$2 / 2$\times$1      &   2   & 1     & 2.3   & 2             & 0.87  & 0\\
2$\times$3 / 3$\times$2      &   6   & 3     & 6.25  & 5             & 0.96  & 0\\
3$\times$3                   &   9   & 5     & 8.9   & 6             & 1.011 & 0\\
2$\times${$k$} / {$k$}$\times$2 \cite{ParandehAfshar11}      &  2$k$ & $k$+1 & 1.65$k$ + 2.3 & $k$+2 & $\frac{2k}{\text{cost}^\text{tile}_t}$    & 0\\
\cmidrule(rl){1-7}
   24$\times$17 / 24$\times$17  &  408  & 0     & 26.65          & 41& 15.30 &1\\
\bottomrule
\end{tabular}
\end{adjustbox}
\label{table:tiles_properties}
\end{table}

\section{Booth-Array Multiplication}
The concept of higher radix multiplication like radix-4 as in the utilized Booth-Array implementation is motivated by the quest to reduce the number of partial products and size of the compression circuits to overall reduce the complexity of the multiplication operation. But when calculating a radix-4 $w_x\times w_y$ multiplication with a digit-set of $X_i \in \{0,1,2,3\}$
\begin{equation}
    P=X\times Y= \left(\sum_{i=0}^{\frac{w_x}{2-1}}4^iX_i\right)\times Y ,
\end{equation}
the multiplication by 3 is computationally expensive to realize compared to the multiplication by 1 and 2, as it would require an additional adder in the critical path, negating the radix-4 benefits of effectively halving the number of partial product bits. To resolve this issue the Booth-Encoding was devised \cite{Booth51} which is using a redundant number system digit set $X_i \in \{-2,-1,0,1,2\}$ for one of the multiplicands and thereby avoids the undesirable multiplication by 3. The Booth-Encoding is performed according to Table~\ref{table:booth_encoding}, such that according to the individual bits $y_{m+1}, y_m, y_{m-1}$ of the multiplicand $Y$ processed in a particular column of the Booth-Array three flags $z_m, c_m$ and $s_m$ are calculated. Those bits control the behavior of the subsequent partial product calculation with the particular bit $x_n$ of the other multiplicand $X$. 
First, there is the zero bit $z_m$ indicating that the result of the booth encoding is $0$. 
Additionally, there is the complement bit $c_m$, denoting a negative sign of the Booth-Encoders output value and the shift bit $s_m$ to represent the absolute value of the booth encoder being $2$.   

\begin{table}[t]
\caption{Truth table for the Booth encoding \cite{Kumm15}}
\centering
\begin{tabular}{ c c c c c c c c}
\toprule
$y_{m+1}$&$y_m$&$y_{m-1}$&$BE_m$&$z_m$&$c_m$&$s_m$\\
\midrule
0&0&0&0&1&0&0\\
0&0&1&1&0&0&0\\
0&1&0&1&0&0&0\\
0&1&1&2&0&0&1\\
1&0&0&-2&0&1&1\\
1&0&1&-1&0&1&0\\
1&1&0&-1&0&1&0\\
1&1&1&0&1&1&0\\
\bottomrule
\end{tabular}
\label{table:booth_encoding}
\end{table}

The radix-4 Booth-Array is well suited for the 6-input LUTs and fast-carry-chains of most modern FPGAs, while higher radix multipliers usually do not map efficiently, as they typically require more than 6 inputs and reintroduce the problem with the multiplication by $3$. 

\begin{figure*}[ht!]
\centering
\begin{subfigure}[b]{0.48\textwidth}
    \centering
    \includegraphics[width=1\textwidth]{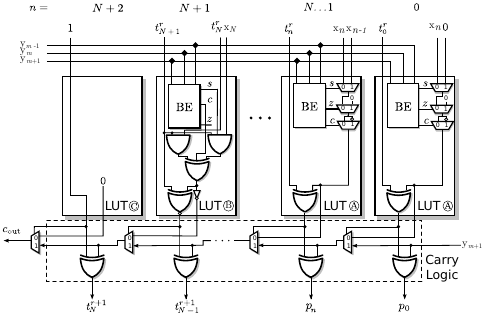}
    \caption{unsigned}
    \label{fig:booth_slice_unsigned}
\end{subfigure}
\begin{subfigure}[b]{0.48\textwidth}
    \centering
    \includegraphics[width=1\textwidth]{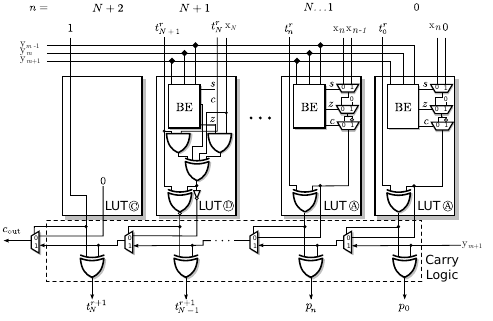}
    \caption{signed}
    \label{fig:booth_slice_signed}
\end{subfigure}
\caption{Slice configuration for Booth multipliers}
\label{fig:booth_slice}
\end{figure*}
\begin{figure*}[ht!]
\centering
\begin{subfigure}[b]{0.62\textwidth}
    \centering
    \includegraphics[width=1\textwidth]{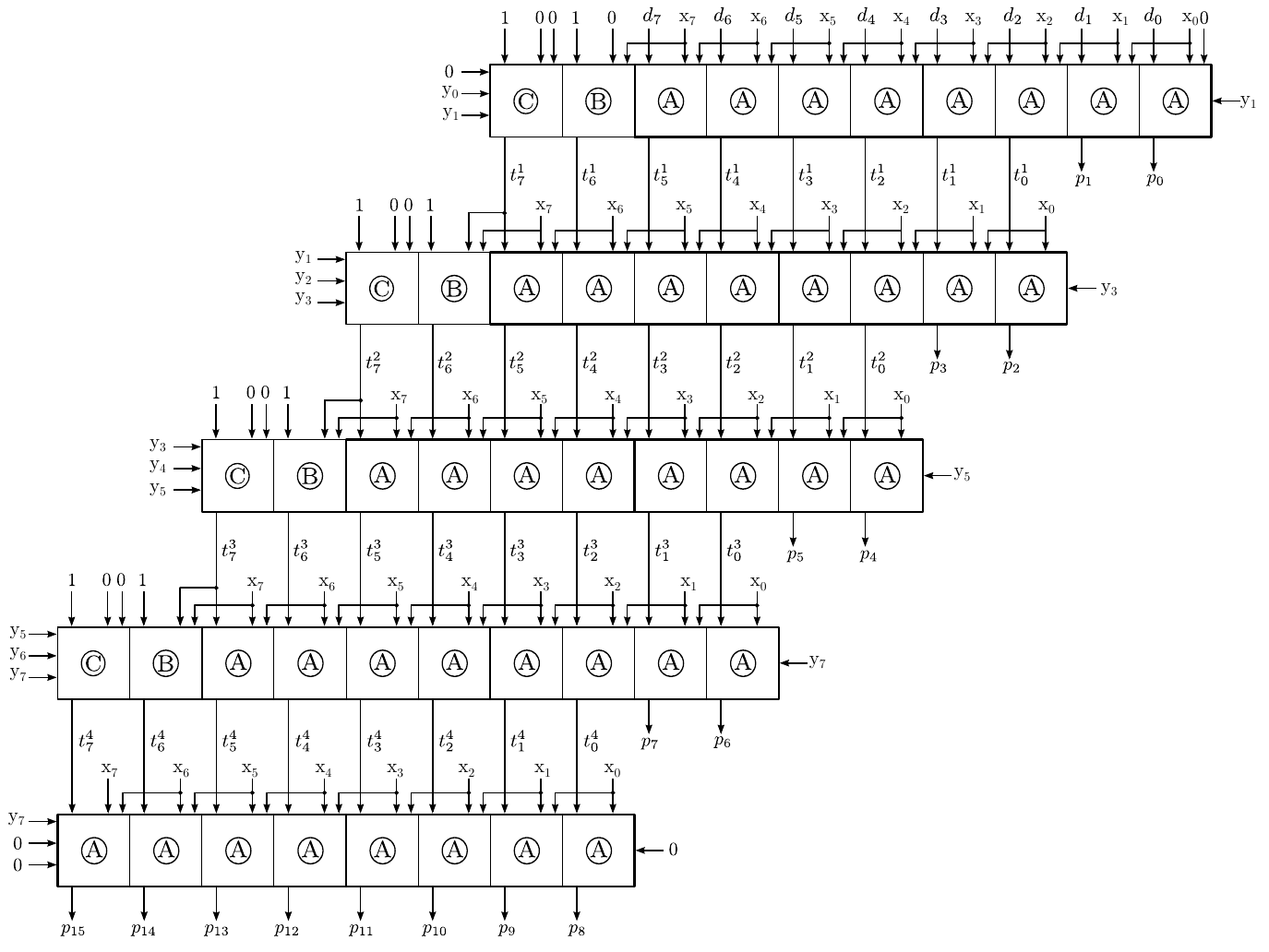}
    \caption{unsigned}
    \label{fig:booth_overall_unsigned}
\end{subfigure}
\hspace{-5cm}
\begin{subfigure}[b]{0.62\textwidth}
    \centering
    \includegraphics[width=1\textwidth]{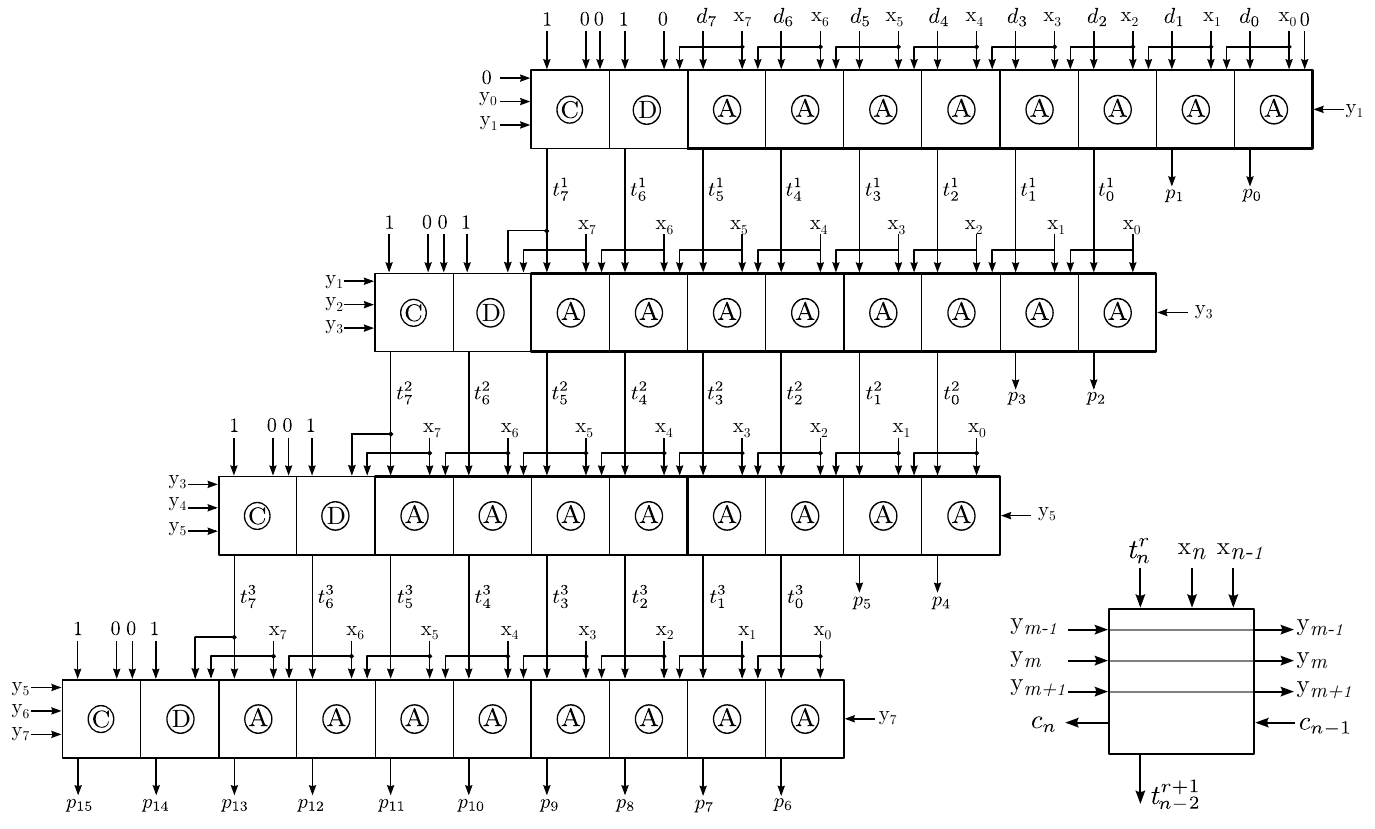}
    \caption{signed}
    \label{fig:booth_overall_signed}
\end{subfigure}
\caption{Overall Structure of a $8\times8$ Booth multiplier}
\label{fig:booth_overall}
\end{figure*}

\figurename~\ref{fig:booth_slice} shows the LUT mapping and Slice configuration for singed and unsigned Booth-Arrays on AMD 7-series \& Ultrascale FPGAs. The considered design incorporates the handling of the signed multiplication of \cite{Bewick94} as implemented structurally in \cite{Walters16,Kumm15}. Each logic-level within the overall architecture in \figurename~\ref{fig:booth_overall} is realized along the fast-carry-chain, such that there is one LUT to process each bit $x_n$ of the $X$-input and its relative LSB $x_{n-1}$. Those share a common LUT-mapping denoted as Type \circled{A} in \figurename~\ref{fig:booth_slice} and contain a Booth encoder (BE) to control the processing of $x_n$ and $x_{n-1}$ depending on $y_{m+1}, y_m, y_{m-1}$. Every LUT generates a partial product bit $t_n^{r+1}$ and adds a partial product bit from the previous logic level $t_n^r$, which result is sent to the subsequent level or constitutes the outputs of the circuit for the two LSBs of each level and the final level. Additionally at position $N+1$ and $N+2$ a modified mapping \circled{B},\circled{C} and \circled{D} is required to handle the sign extension and carries from the lower bits. 

The overall structure of the Booth-Array is shown in \figurename~\ref{fig:booth_overall} on the example of an $8\times 8$ multiplier. This also illustrates the major drawback of Booth multipliers as with multiple stacked logic levels, the critical path delay (or latency when pipelined) quickly rises with (vertical) size, as the partial products from earlier stages are compressed by the respective subsequent stages via the $t_n^r$ signals. The delays for routing between the logic levels can be over-proportionally long compared to the delay along the fast carry chain, which is unusually can be placed in physical proximity for adjacent slices, while the routed design typically includes a considerably slower, more convoluted path between the slices of different logic levels. Since also the first booth-level uses the A-Type mapping which features a compressor input $t_n^r$, as suggested in \cite{Kumm15}, the Booth-Array can be used to calculate a MAC-operation, with the same width $N$ as input $X$ without additional costs. These accumulate inputs $d_n$ have to be set to $0$ when this feature is unused.
One of the major structural differences between the singed and unsigned version is that in the signed cases of the $8\times 8$-multiplier only 4 logic levels are required (\figurename~\ref{fig:booth_overall_signed}) when applying the optimizations of \cite{Bewick94, Walters14, Walters16}, while the unsigned case requires 5 levels (\figurename~\ref{fig:booth_overall_unsigned}) and only the 2 MSB LUTs in the final level can be saved. Accordingly the unsigned Booth-Array requires one level more in sizes with even height or conversely the height of the signed case can be one bit more with almost the same cost.

\section{Integration of Booth-Arrays into Multiplier Tiling}
As aforementioned practically all multiplier designs can also be used as sub-multiplier for tiling. Aside from the availability of the implementation, the tile has to be abstracted for the tiling algorithm with the properties described in Table~\ref{table:tiles_properties}.

\subsection{Evaluation of Cost Relative to Booth-Level}
The utilized multiplier tiling optimization model \cite{Boettcher23} solely bases the decision to place a particular tile at a certain position on the overall realization cost of the resulting multiplier and the specified constraints in terms of DSP usage. So a sensible pre-selection considering for other factors like critical path delay has to be performed beforehand to decide which sub-multipliers to include in the tile-set. The properties of Booth-Arrays with different logic-depths are shown in Table~\ref{table:booth_level_eff} in terms of the metrics introduced with Table~\ref{table:tiles_properties}.
\begin{figure}[h]
    \centering
    \includegraphics[width=0.46\textwidth]{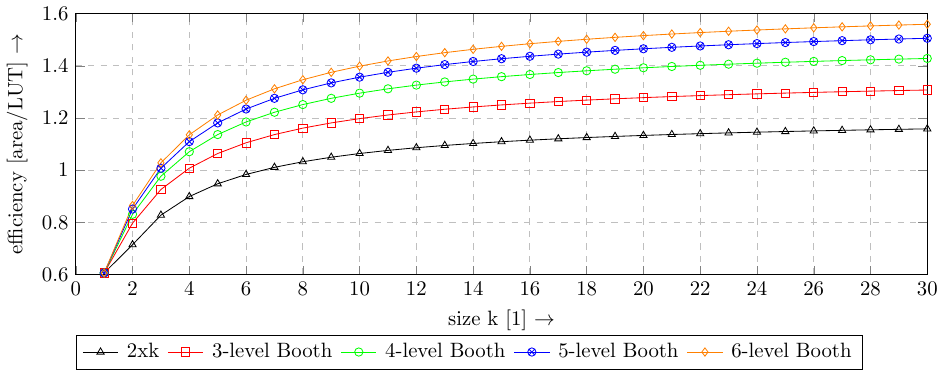}
    \caption{Dependency between size and efficiency}
    \label{fig:eff_vs_size}
\end{figure}
Additionally, \figurename~\ref{fig:eff_vs_size} shows the efficiency in terms of area per LUT for Booth multipliers with different logic levels and the 2$\times${$k$}-multiplier for reference. It can be seen that the efficiency of the evaluated designs increases with size, as there is some overhead for the basic structure of the architecture which relativizes with size $k$. For the 2$\times${$k$}-multiplier that is 1 LUT, while it is 2 LUTs per level for the Booth multiplier except the unsigned final level. Booth-Arrays are significantly more efficient than the 2$\times${$k$}-multiplier which has a peak efficiency of 
$\lim \limits_{k \to \infty}\frac{2}{1.65+\frac{2.3}{k}}=1.21$, compared to the different level of Booth multipliers in Table \ref{table:booth_level_eff}. Also note that there are diminishing returns in terms of efficiency increase for each additional logic level which conversely contributes to the critical path delay. 

\begin{table}[t]
\caption{Properties of different Level Booth Multiplier 
on AMD FPGAs, where $\text{cost}^\text{mult}_e$ and $\text{cost}^\text{tile}_t$ are the number of LUT6 for the single multiplier and multiplier+compression, respectively
}
\centering
\begin{adjustbox}{max width=0.48\textwidth}
\setlength\tabcolsep{3pt}
\begin{tabular}{ l c c c c c c c}
\toprule
Dimension & $A_t$ & $\text{cost}^\text{mult}_t$ & $\text{cost}^\text{tile}_t$ & $w_\text{out}$ & $E_t$ & $\!\!\lim \limits_{k \to \infty}\!\!E_t(k)$\\
\midrule
5$\times$k / {$k$}$\times$5     &  5$k$ & $3\left(k\!+\!1\right)$  & $6.25+3.65k$ & $k$+5 & $\frac{5k}{6.25+3.65k}$ & 1.37\\
7$\times$k / {$k$}$\times$7     &  7$k$ & $4\left(k\!+\!1\right)$  & $8.55+4.65k$ & $k$+7 & $\frac{7k}{8.55+4.65k}$ & 1.51 \\
9$\times$k / {$k$}$\times$9     &  9$k$ & $5\left(k\!+\!1\right)$  & $10.85+5.65k$ & $k$+9 & $\frac{9k}{10.85+5.65k}$ & 1.59 \\
11$\times$k / {$k$}$\times$11   &  11$k$ & $6\left(k\!+\!1\right)$  & $13.15+6.65k$ & $k$+11 & $\frac{11k}{13.15+6.65k}$ & 1.65 \\
\bottomrule
\end{tabular}
\end{adjustbox}
\label{table:booth_level_eff}
\end{table}

\subsection{Evaluation of Delay Relative to Booth Level}
Every additional level of the Booth-Array introduces an additional delay to the circuit, so when delay (or latency) is considered as a secondary objective, a trade-of-point has to be found, to avoid the partial product-generation with the Booth-Array to contribute over-proportionally to the total delay of the design. To evaluate the relation between the Booth-level and the critical path delay a series of synthesis experiments was performed for Booth multipliers with a fixed width of $w_X=32$ and different number of levels (3,\ldots,6).
\begin{table}[t]
\caption{Synthesis experiments for different Booth levels}
\centering
\begin{adjustbox}{max width=0.48\textwidth}
\setlength\tabcolsep{3pt}
\begin{tabular}{ l c c c c c c c}
\toprule
 & \multicolumn{3}{c}{\textbf{unsigned}} & \multicolumn{3}{c}{\textbf{signed}}\\
 \cmidrule(rl){2-4} \cmidrule(rl){5-7} 
Level& Size& LUTs&CPD [ns]\! & Size&  LUTs&CPD [ns]\! \\
\midrule
3&5$\times${$k$}    &  99&4.2 & 6$\times${$k$}    &99&4.2 \\
4&7$\times${$k$}    & 132&4.9 & 8$\times${$k$}    &132&4.9  \\
5&9$\times${$k$}    & 165&6.5 &10$\times${$k$}    &165&6.9\\
6&11$\times${$k$}   & 198&7.7 &12$\times${$k$}    &198&8.0\\
\bottomrule
\end{tabular}
\end{adjustbox}
\label{table:booth_level_cpd}
\end{table}
The experiments were performed with signed and unsigned Booth-Arrays with the maximal vertical size with a particular level, which is one bit more for the signed case. 
The results are shown in Table~\ref{table:booth_level_cpd}.
When the maximal vertical size for a particular level is utilized the resulting structure is almost exactly the same, except the mapping of LUT \circled{B} or \circled{D}, which explains the equal resource costs, in contrast to \figurename~\ref{fig:booth_overall_unsigned} where the array is effectively under-utilized. In terms of the resulting critical path delay (CPD), an over-proportional increase can be observed between level 4 and 5 in Table~\ref{table:booth_level_cpd}. 
Additionally, as it is evident in \figurename~\ref{fig:eff_vs_size}, increasing the Booth levels beyond 4 results in a considerably smaller increase in efficiency than from 3 to 4 levels.
So, it was decided to use Booth-arrays of up to 4 levels as sub-multipliers for tiling as this constitutes a good compromise taking both resources and delay into account.

\section{Results}
\subsection{Experimental Setup}
To generate results for multiplier tiling, the combined optimization of tiling, compressor tree and the final adder of \cite{Boettcher23} was used as a state-of-the-art tiling method.
For that, the open-source arithmetic core generator FloPoCo \cite{dk24} was extended to support the Booth-Array multipliers as tiles.
The results of this work are made open-source as well and are available as part of the \texttt{IntMultiplier} operator in the current FloPoCo git\footnote{see \url{https://www.flopoco.org}}.
The combined tiling and compression method uses integer linear programming (ILP) to generate results which are optimal for a given set of tiles, compressors and final adder architectures. Gurobi 11.0 is used as the ILP-Solver with a timeout of 10h on a computer with an Intel i9-12900K with 128GiB RAM running Linux Mint 21.3. The synthesis experiments were performed with Vivado 2022.1 for an AMD Kintex 7 (xc7k70tfbv484-3) target FPGA. The timing data were obtained by synthesizing with a register sandwich.
\subsection{Evaluation of the Impact of the Introduction of Booth-Arrays to Tiling}
Multiplier tiling traditionally only used DSPs and logic-based tiles with only one level of logic, which ensures a low critical path delay and latency when pipelined. Introducing sub-multipliers with multiple levels like the Booth-Array as tiles is expected to cause a certain increase to the critical path delay, while the lower costs per area can reduce the overall realization cost of the designed multipliers.

\begin{figure*}[]
\centering
\begin{subfigure}[b]{0.46\textwidth}
    \centering
    \includegraphics[width=1\textwidth]{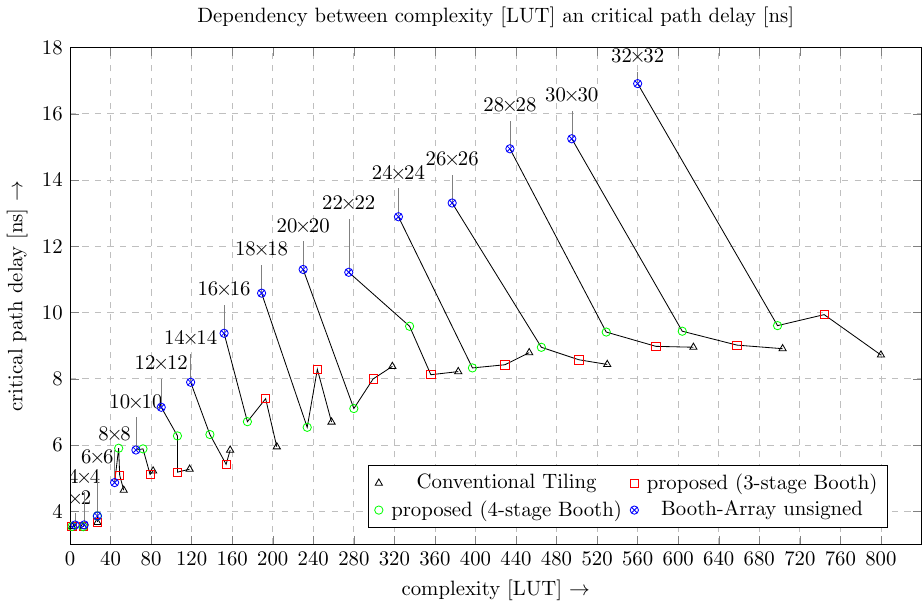}
    \caption{unsigned, 0DSP, not pipelined}
    \label{fig:lut_vs_cpd_s0_d0_p0}
\end{subfigure}
\hspace{1cm}
\begin{subfigure}[b]{0.46\textwidth}
    \centering
    \includegraphics[width=1\textwidth]{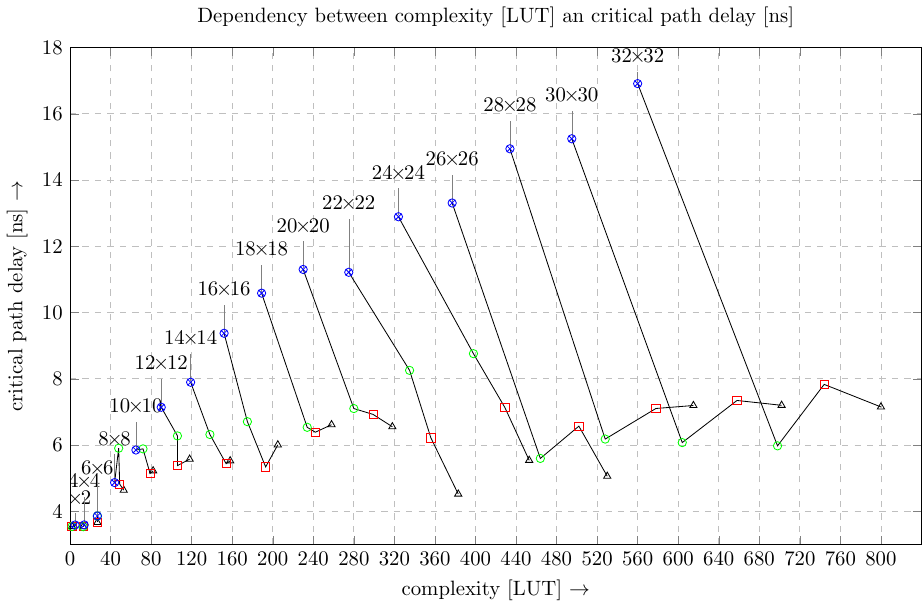}
    \caption{unsigned, 0DSP, pipelined}
    \label{fig:lut_vs_cpd_s0_d0_p1}
\end{subfigure}

\begin{subfigure}[b]{0.46\textwidth}
    \centering
    \includegraphics[width=1\textwidth]{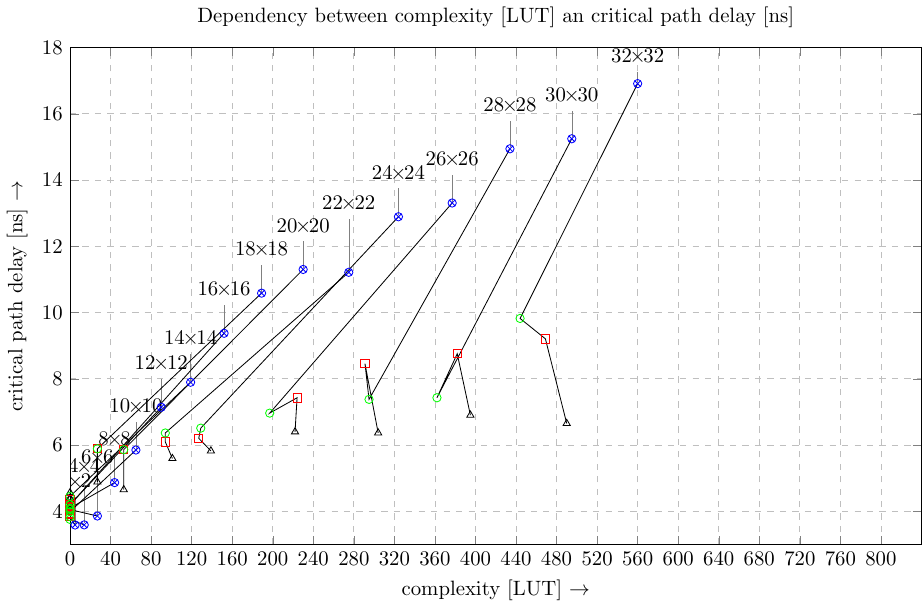}
    \caption{unsigned, 1DSP, not pipelined}
    \label{fig:lut_vs_cpd_s0_d1_p0}
\end{subfigure}
\hspace{1cm}
\begin{subfigure}[b]{0.46\textwidth}
    \centering
    \includegraphics[width=1\textwidth]{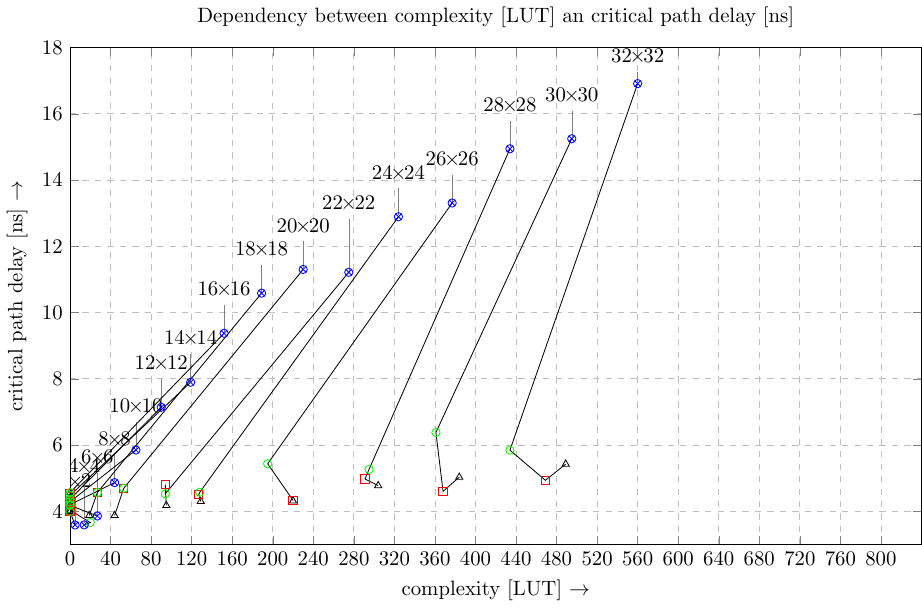}
    \caption{unsigned, 1DSP, pipelined}
    \label{fig:lut_vs_cpd_s0_d1_p1}
\end{subfigure}

\begin{subfigure}[b]{0.46\textwidth}
    \centering
    \includegraphics[width=1\textwidth]{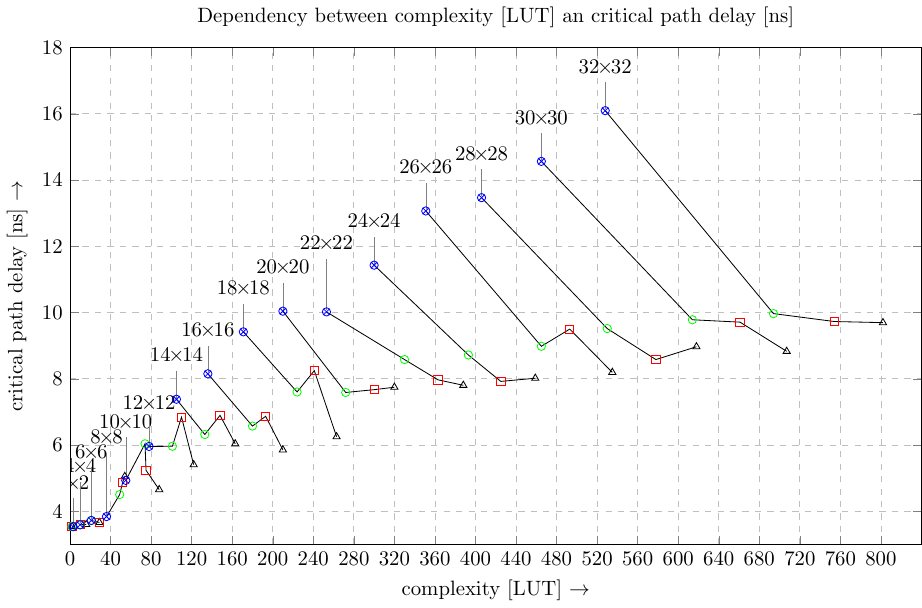}
    \caption{signed, 0DSP, not pipelined}
    \label{fig:lut_vs_cpd_s1_d0_p0}
\end{subfigure}
\hspace{1cm}
\begin{subfigure}[b]{0.46\textwidth}
    \centering
    \includegraphics[width=1\textwidth]{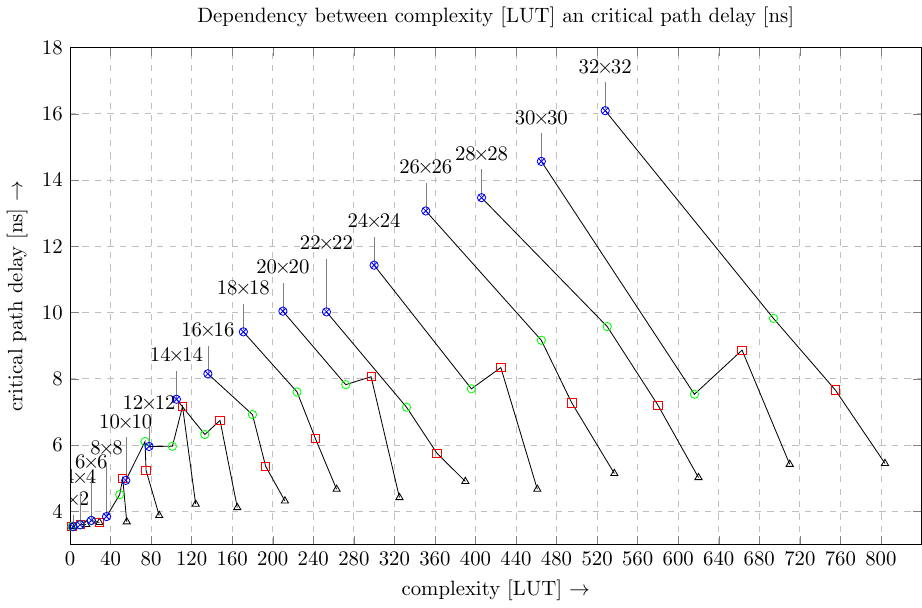}
    \caption{signed, 0DSP, pipelined}
    \label{fig:lut_vs_cpd_s1_d0_p1}
\end{subfigure}

\begin{subfigure}[b]{0.46\textwidth}
    \centering
    \includegraphics[width=1\textwidth]{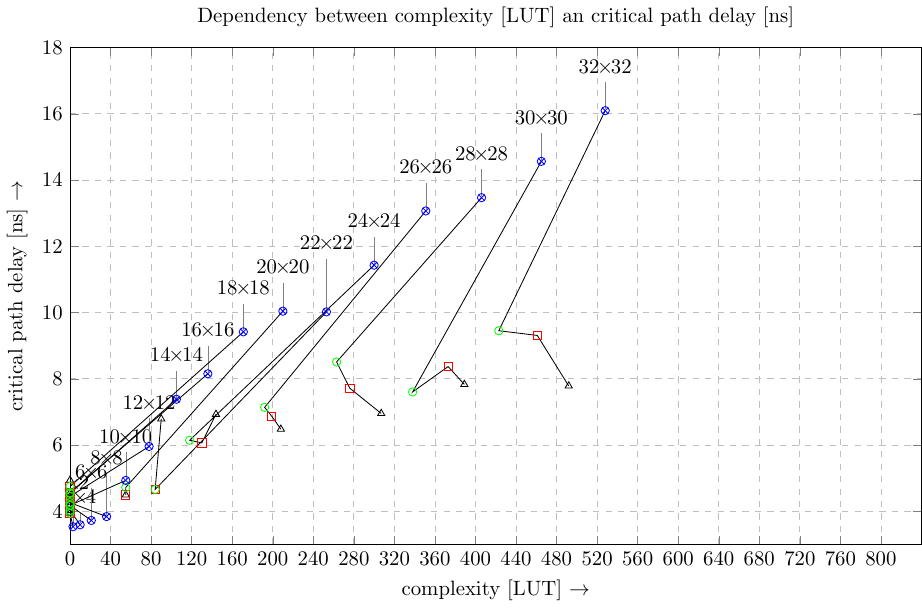}
    \caption{signed, 1DSP, pipelined}
    \label{fig:lut_vs_cpd_s1_d1_p0}
\end{subfigure}
\hspace{1cm}
\begin{subfigure}[b]{0.46\textwidth}
    \centering
    \includegraphics[width=1\textwidth]{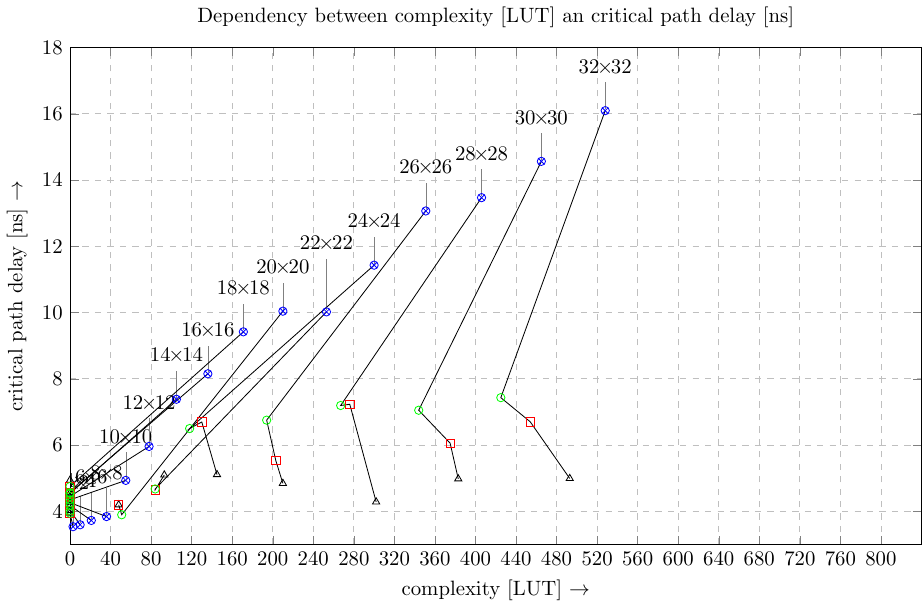}
    \caption{signed, 1DSP, pipelined}
    \label{fig:lut_vs_cpd_s1_d1_p1}
\end{subfigure}
\caption{critical path delay vs. complexity}
\label{fig:cpd_vs_complex}
\end{figure*}

To evaluate the impact, synthesis experiments featuring tiling with the previous tile-set, tiling with 3- and 4-level Booth-Arrays and a full sized Booth multiplier were performed for multipliers from 2$\times$2 to 32$\times$32 bit, with 0 or 1 DSP, pipelined or combinatorial and signed or unsigned cases. 
Pipelining resulted in one and two stages for sizes of larger than about $8\times 8$ and $30\times 30$, respectively.
All designs up to about $22\times 22$ were solved optimally by the ILP solver, for larger multipliers there have been cases where the best feasible result within the timeout was taken. 
The results are visualized in figs.~\ref{fig:cpd_vs_complex} plotting the critical path delay over the complexity in LUT. 
Note that the large Booth-Array in figs.~\ref{fig:lut_vs_cpd_s0_d0_p1},\ref{fig:lut_vs_cpd_s0_d1_p1},\ref{fig:lut_vs_cpd_s1_d0_p1},\ref{fig:lut_vs_cpd_s1_d1_p1} is for reference and itself not pipelined. 
It can be seen, that as expected, in general the tilings without Booth-Arrays tend to be the fastest, followed by the 3- and 4-level and the full-size combinatorial Booth-Array.
 In terms of resources the opposite holds true, so the previous tiling is the most costly followed by the tilings with 3- and 4-level Booth sub-multipliers, with the large Booth-Array being typically the least expensive variant. 
 As expected, when 1 DSP is permitted within the tiling (figs.~\ref{fig:lut_vs_cpd_s0_d1_p0},\ref{fig:lut_vs_cpd_s0_d1_p1},\ref{fig:lut_vs_cpd_s1_d1_p1},\ref{fig:lut_vs_cpd_s1_d1_p0}), the tiling-based multipliers are usually less expensive than the large Booth multiplier. 
Overall, we observe a consistent generation of new Pareto optimal points, offering a significant reduction of resources by introducing a slight delay increase when considering Booth multipliers in the tiling.

\subsection{Comparison to State-of-the-Art Designs}

\begin{table}[]
\caption{Comparison to State-of-the-Art Designs}
\centering
\begin{adjustbox}{max width=0.48\textwidth}
\begin{tabular}{L{0.1cm} L{0.1cm}  L{0.1cm} L{2.3cm} C{0.2cm} C{0.2cm} C{0.2cm} C{0.2cm} C{0.2cm} C{0.2cm} C{0.3cm} C{0.3cm} C{0.3cm} C{0.6cm} C{0.4cm} C{0.4cm} }
\toprule
& & &\textbf{Type} & \multicolumn{3}{c}{\textbf{4$\times$4}} & \multicolumn{3}{c}{\textbf{8$\times$8}} & \multicolumn{3}{c}{\textbf{16$\times$16}} & \multicolumn{3}{c}{\textbf{32$\times$32}}\\
\cmidrule(rl){5-7} \cmidrule(rl){8-10} \cmidrule(rl){11-13} \cmidrule(rl){14-16}\\
&  &  &  &\rotatebox{90}{\textbf{\#LUTs}\!} & \rotatebox{90}{\textbf{CPD [ns]}} & \rotatebox{90}{\textbf{LAT [cyc]}} & \rotatebox{90}{\textbf{\#LUTs}\!} & \rotatebox{90}{\textbf{CPD [ns]}} & \rotatebox{90}{\textbf{LAT [cyc]}} & \rotatebox{90}{\textbf{\#LUTs}\!} & \rotatebox{90}{\textbf{CPD [ns]}} & \rotatebox{90}{\textbf{LAT [cyc]}} & \rotatebox{90}{\textbf{\#LUTs}\!} & \rotatebox{90}{\textbf{CPD [ns]}} & \rotatebox{90}{\textbf{LAT [cyc]}} \\
\midrule
\multirow{20}{*}{\rotatebox{90}{\bf unsigned} } & \multirow{10}{*}{\rotatebox{90}{\bf logic only} } & \multirow{5}{*}{\rotatebox{90}{\bf comb.} }
      &Walters \cite{Walters16}&14&3.6&0&44&4.9&0&152&9.4&0&560&16.9&0\\
&&&Kumm \cite{Kumm15}&17&3.6&0&51&5.2&0&167&8.9&0&591&16.3&0\\
&&&Böttcher \cite{Boettcher23}&13&3.5&0&53&4.6&0&204&6.0&0&800&8.7&0\\
&&&prop. Booth level=3&13&3.5&0&49&5.1&0&193&7.4&0&744&9.9&0\\
&&&prop. Booth level=4&13&3.5&0&48&5.9&0&175&6.7&0&698&9.6&0\\
\cmidrule(rl){3-16}
&&\multirow{4}{*}{\rotatebox{90}{\bf pipelined} }
&Böttcher \cite{Boettcher23}&13&3.5&1&52&3.6&2&199&3.6&3&800&4.0&5\\
&&&Kumm \cite{Kumm15} &19&3.5&3&59&3.6&5&191&3.7&9&647&4.1&17\\
&&&prop. Booth level=3&13&3.6&1&49&3.6&2&193&3.8&3&751&4.0&4\\
&&&prop. Booth level=4&13&3.5&1&42&3.7&3&175&3.8&3&715&4.3&4\\
\cmidrule(rl){2-16}
& \multirow{7}{*}{\rotatebox{90}{\bf 1 DSP} } & \multirow{3}{*}{\rotatebox{90}{\bf comb.} }
&Böttcher \cite{Boettcher23}&0&4.0&0&0&4.2&0&0&4.6&0&490&6.7&0\\
&&&prop. Booth level=3&0&3.9&0&0&4.0&0&0&4.4&0&469&9.2&0\\
&&&prop. Booth level=4&0&3.8&0&0&4.1&0&0&4.2&0&444&9.8&0\\
\cmidrule(rl){3-16}
&&\multirow{3}{*}{\rotatebox{90}{\bf pip.} }
  &Böttcher \cite{Boettcher23}&0&4.0&1&0&4.2&1&0&3.8&2&491&5.1&2\\
&&&prop. Booth level=3&0&4.0&0&0&4.3&0&0&4.6&1&469&7.3&1\\
&&&prop. Booth level=4&20&3.5&1&0&4.2&1&0&3.8&2&434&4.9&2\\
\midrule
\multirow{22}{*}{\rotatebox{90}{\bf signed}} & \multirow{21}{*}{\rotatebox{90}{\bf logic only} } & \multirow{10}{*}{\rotatebox{90}{\bf combinatorial} }
 &Walters \cite{Walters16} &10&3.6&0&36&3.8&0&136&8.2&0&528&16.1&0\\
&&&Ullah \cite{Ullah21} &18&3.6&0&66&3.7&0&243&5.4&0&928&8.2&0\\
&&&Ullah \cite{Ullah18} &22&3.3&0&81&5.2&0&296&7.3&0&1121&9.7&0\\
&&&Kulkarni \cite{Kulkarni11} &20&2.1&0&86&4.9&0&330&6.6&0&1257&8.9&0\\
&&&Rehman \cite{Rehman16} &18&2.2&0&92&5.0&0&404&7.0&0&1512&9.6&0\\
&&&AMD~IP~(speed)~\cite{Xilinx2015}&18&3.1&0&72&4.3&0&280&5.9&0&1089&8.8&0\\
&&&AMD IP (area)~\cite{Xilinx2015} &30&2.9&0&51&6.7&0&231&8.3&0&930&9.3&0\\
&&&Böttcher \cite{Boettcher23}&16&3.6&0&54&5.1&0&210&5.9&0&802&9.7&0\\
&&&prop. Booth level=3&10&3.6&0&50&4.8&0&183&6.6&0&738&9.4&0\\
&&&prop. Booth level=4&10&3.6&0&36&3.8&0&169&6.5&0&689&10.4&0\\
\cmidrule(rl){3-16}
&&\multirow{5}{*}{\rotatebox{90}{\bf pipipelined} }
&Böttcher \cite{Boettcher23}&16&3.5&1&56&3.6&2&212&3.8&3&810&4.2&5\\
&&&Booth Dadda\cite{Nagar24} &23&1.8& -- &74&2.7& -- &270&4.0& -- &1054&5.4& -- \\
&&&Booth Wallace\cite{Nagar24} &21&2.1& -- &77&3.1& -- &288&4.1& -- &1130&5.8& -- \\
&&&prop. Booth level=3&10&3.5&2&50&3.7&2&193&3.8&3&757&4.1&4\\
&&&prop. Booth level=4&10&3.7&1&36&3.6&1&163&4.2&2&690&4.8&2\\
\cmidrule(rl){2-16}
& \multirow{6}{*}{\rotatebox{90}{\bf 1 DSP} } & \multirow{3}{*}{\rotatebox{90}{\bf comb.} }
 &Böttcher \cite{Boettcher23}&0&4.0&0&0&4.3&0&0&4.6&0&492&7.8&0\\
 &&&prop. Booth level=3&0&4.0&0&0&4.3&0&0&4.6&0&446&9.9&0\\
 &&&prop. Booth level=4&0&4.0&0&0&4.3&0&0&4.6&0&409&8.6&0\\
\cmidrule(rl){3-16}
&&\multirow{3}{*}{\rotatebox{90}{\bf pip.} }
&Böttcher \cite{Boettcher23}&0&4.0&0&0&4.3&0&0&4.6&1&493&5.0&2\\
&&&prop. Booth level=3&0&4.0&1&0&4.3&1&0&4.6&1&441&4.9&2\\
&&&prop. Booth level=4&0&4.0&1&0&4.3&1&0&3.8&2&409&5.1&2\\
\bottomrule
\end{tabular}
\end{adjustbox}
\label{table:sota_comparison}
\end{table}

To evaluate the properties of the proposal in terms of LUT-resources and timing (critical path delay (CPD) [ns], latency (LAT) in cycles) in comparison to the state-of-the-art, synthesis experiments were conducted for various test-cases of signed and unsigned multipliers.
Whenever reference implementations were available, the syntheses experiments were repeated for the reference designs to ensure equal test conditions. 
As expected, the unpipelined Booth-Arrays \cite{Kumm15,Walters16} in Table~\ref{table:sota_comparison} have the longest critical path delay, while resource-wise achieving one of the best results without the use of DSPs. As for the previous tiling \cite{Boettcher23}, the proposal allows to offset complexity from LUT-resources to DSPs, when one DSP is permitted within the design, the LUT-costs are lower than the reference designs, or are 0 when the smaller designs solely consists of the DSP. With regards to the timing, the use of DSPs helps to reduce the critical path delay proportionally the reduction of the logic based part of the design, as the router has to place fewer paths which reduces the likelihood for one path to introduce an excessive delay. Surprisingly, the AMD multiplier IP-core optimized for speed is less costly than the area optimized version while still being faster. As expected, the previous tiling \cite{Boettcher23} tends to be slightly faster than the proposal. The tiling with 4-level Booth is slightly slower than the 3-level variant with the magnitude expected from Table~\ref{table:booth_level_cpd}. The previous proposals to mitigate the long critical path delay of a combinatorial Booth-Array \cite{Kulkarni11,Rehman16,Ullah18,Ullah21,Nagar24} are effective in their goal, although at a considerable expense of logic resources, due to the realization cost of the compressor tree. The proposal generates structurally similar designs, but with a more efficient compressor tree, which results in fewer compressor stages and an overall less expensive design. 
With respect to timing, similar values are achieved for the critical path delay (CPD) when realized as purely combinatorial circuits (latency (LAT) of  0 cycles). In \cite{Nagar24} it remains unclear if the designs are pipelined, as they have similar structure and levels of the compressor tree as the main contributing factor to the delay as other reference designs, but the critical path delay corresponds to typical values for a lower number of stacked logic levels. 
Due to their CPD, we considered them as being pipelined.
In general similar timing results than other architectures that seek to reduce the delay of Booth-Arrays can be achieved under comparable conditions, while typically achieving considerable resource reductions. When faster designs are needed, the delay can effectively be further reduced by pipelining at the expense of latency.

\section{Conclusion}
The present work demonstrates a systematic approach to address the inherent disadvantage of Booth-Arrays, that the stacking of logic levels introduces a considerable delay in the design, by assembling a large multiplier from smaller Booth-Arrays and other sub-multipliers. The partial products are then subsequently compressed on a compressor tree with state-of-the-art compressors, which results in fewer stacked logic levels than within a monolithic Booth-Array with the same size or other recent techniques to reduce the delay of the Booth-Array. This in turn results in cost reductions compared to other mitigation techniques, while similarly reducing the delay. It is shown that using Booth-Arrays up to 4 levels as sub-multipliers constitutes a sensible compromise between cost and delay. The use of DSPs alongside Booth-Arrays in multiplier-tiling is demonstrated and shown to be beneficial with regards to cost and delay. It is demonstrated, that optimal solutions for a given set of compressors and tiles including Booth-Arrays can be found with a previous ILP-model for the combined global optimization of tiling and compressor tree design.

\balance

\bibliographystyle{IEEEtran}
\bibliography{IEEEabrv,bibliography}

\end{document}